\documentclass[12pt]{amsart}
\usepackage[utf8]{inputenc}
\usepackage{pgfplots}

\def\Re{\mathbf{R}}

\def\ta{\theta}
\def\al{\alpha}
\def\la{\lambda}
\def\da{\delta}
 
\def\phi{\varphi}

\newcommand{\df}[1]{\textit{#1}}
\newcommand{\norm}[1]{\| #1 \|}

\def\ngoods{n}

\usepackage{natbib}

	\usepackage[left=1.5in,top=1.3in,right=1.5in,bottom=1.3in]{geometry}
	\usepackage[onehalfspacing]{setspace}

\newtheorem*{proposition*}{Proposition}
\newtheorem*{obs}{Observation}

\title[On the CCEI]{On the meaning of the \\ Critical Cost Efficiency Index}
\author[Echenique]{Federico Echenique}
\date{April 2022}
\thanks{\textit{Affiliation: } Division of the Humanities and Social Sciences, California Institute of Technology. 
I thank Roy Allen, Laurens Cherchye, Bram De Rock, Thomas Demuynck, Pawe\l\ \citeauthor{DZIEWULSKI2020105071}, Yoram Halevy, 
Taisuke Imai, Shachar Kariv, SangMok Lee, Matt Polisson, Matt Shum, and John Quah for discussing this note with me. This does not mean that they agree with the views I express here. I am also grateful to the National Science Foundation for financial support through grant SES-1558757.}

\begin{document}

\begin{abstract}
    This note provides a critical discussion of the \textit{Critical Cost-Efficiency Index} (CCEI) as used to assess deviations from utility-maximizing behavior. I argue that the CCEI is hard to interpret, and that it can disagree with other plausible measures of ``irrational'' behavior. The common interpretation of CCEI as wasted income is questionable. Moreover, I show that one agent may have more unstable preferences than another, but seem more rational according to the CCEI. This calls into question the (now common) use of CCEI as an ordinal and cardinal measure of degrees of rationality.
\end{abstract}

\maketitle

\section{Introduction}

Afriat's \textit{Critical Cost Efficiency Index} (CCEI) has become the standard measure of ``irrational'' behavior in many empirical studies using both field and laboratory data. A search on Google Scholar reveals at least 70 papers with empirical studies that use the CCEI, including highly-cited publications in leading journals.  Most of the papers are recent, meaning that the methodology is gaining popularity.

I believe that the CCEI has drawbacks that do not seem to be widely known or understood: existing empirical work using it does not qualify their conclusions, or include any critical methodological discussion. I mean to articulate these drawbacks to further a better understanding of the use of CCEI. In doing so, I will avoid the critique of any individual paper. The CCEI is very popular, and 
the authors of the work I have surveyed have been following what is now considered standard practice by the profession. Use of the CCEI is widespread, and by being associated to one of the most famous ideas in revealed preference analysis (Afriat's beautiful study of rationalizable consumer choices), it has impeccable intellectual pedigree.  With this note, I hope to make practitioners become more aware of the potential problems involved in using the index, and spark more research into alternative measures.

The empirical work I have reviewed uses the CCEI for three different purposes. The first is to decide whether a dataset exhibits a  violation of ``rationality,'' meaning a violation of the utility-maximization hypothesis. When a dataset is not rationalizable, CCEI is used to decide whether the data warrants a rejection of the utility-maximization hypothesis. The second purpose is to compare different datasets that violate rationality, and understand if one is more irrational than the other. The data usually corresponds to different individual agents, and the question is who is more rational than whom. The third use is as a cardinal measure of irrationality. I shall argue that all three uses of CCEI are problematic.

The rest of this introduction assumes familiarity with revealed preference theory, the CCEI, and the generalized axiom of revealed preference (GARP). The setting is that of textbook demand theory, in which a single agent chooses consumption bundles at a given price and budget. A dataset consists of a finite collection of observations, drawn from different prices and budgets. The utility maximization hypothesis is captured by a property of the data, the absence of revealed-preference cycles, which is termed GARP (the ``generalized axiom of revealed preference''). Such cycles may be avoided by shrinking observed budgets for the purposes of making revealed preference comparisons, and the CCEI is the least shrinking needed for GARP to be satisfied. The beginning of Section~\ref{sec:modelandexample} provides formal definitions and a brief explanation of the theory (see \cite{chambers2016revealed} for an in-depth exposition of revealed preference theory).

\subsection{Interpretation of the CCEI}

The first and most basic problem with CCEI is its interpretation. Say that the data from one consumer return a CCEI of 0.8. What does that number mean? This question speaks to the first kind of application of CCEI, when we want to decide whether a given measurement for the index warrants the conclusion that the consumer is irrational. Is 0.8 good or bad?

Most empirical papers interpret a value of 0.8 as the consumer ``wasting'' 20\% of their income, or leaving 20\% of their income on the table. Such language is appealing because it appears to capture irrationality in stark monetary terms; and because the notion of waste strongly suggests irrational behavior. The exact story, though, is never quite pinned down. The papers that use the CCEI do not walk the reader through an explanation of how the number 0.8 connects to a 20\% waste.

Logically, the waste interpretation depends on the existence of some measure of welfare that can be achieved more cheaply than what was observed in the data. The consumer with $\textrm{CCEI}=0.8$ could have achieved the same level of well-being with 20\% less income. Even if the assumption of the existence of a measure of welfare is not spelled out, it seems unavoidable if one is to reach the waste conclusion. How else could we say that there is waste if there is not some objective that is achieved, or outcome that is preserved, with less money? Unfortunately the exact measure of welfare is never clarified. Clearly, the welfare measure cannot be utility, as no utility is compatible with the consumer's choices. \cite{afria72}, in fact, seemed to recognize the difficulties inherent in consumption data, where there is no observable measure of ``output.'' 
\footnote{\cite{afria72} writes:
``In the corresponding approach to demand
 analysis already elaborated, the algebraical development is different because the hypothetical output of consumption has a different metrical character, being
 without a specific magnitude and hence also without a price.'' In his formulation of the CCEI for production (actually cost minimization), efficiency is relative to a target output level. In the quote, he seems to point out that there is no equivalent metric in the case of consumption data. And when he refers to an analysis ``already elaborated'' he seems to mean \cite{afriat1973system}, where he proposes the CCEI.} If not utility, what could welfare be?

Once utility is ruled out, it is difficult to envision alternative welfare objectives that could justify the waste interpretation. One idea is that the observed bundles can be purchased with less income. Consider an example with two goods and a dataset with two observations: $(x^1,p^1)=((4/10,8/10),(1/2,1))$ and $(x^2,p^2)=((1,0),(1,1/2))$. Expenditure equals $1$ in each observation, while $p^1\cdot x^2=1/2$ and $p^2\cdot x^1=0.8$. It is easy to verify that $\textrm{CCEI}=0.8$, and indeed one might imagine the consumer purchasing bundles $x^1$ and $x^2$ at prices $p^2$, for a total expenditure of $1.8$ rather than $2$. This would seem to indicate a waste of $0.2$, in line with the common interpretation of CCEI: The same bundles are purchased with $0.2$ fewer units of money. At the same time, the consumer could just as reasonably purchase $x^1$ at prices $p^2$ and $x^2$ at prices $p^1$, for a total expenditure of $1.3$. Now we have saved much more, and the ``waste'' would come to 35\%. So I do not see how waste connects to the CCEI, even when the welfare objective is reduced to maintaining the same bundles as in the data.\footnote{Actually this interpretation of waste would seem to lead to the money pump index of \cite*{echenleeshum2011}. This example might seem directed at a straw man, but addresses a comment I have received in response to the note.}

\cite{afriat1973system} did not describe CCEI as measuring waste, but rather as the ``cost efficiency'' at which the agent would be optimizing. This interpretation is laid out with great clarity in the recent note by  \cite{polisson2021rationalizability}:  A utility function $u$  has cost efficiency $0.8$ for bundle $x$ at prices $p$ if the consumer cannot surpass a utility of $u(x)$ and at the same time spend 20\% less than the budget $p\cdot x$.\footnote{See also Theorem 2 in \cite*{halevy2018parametric}.} One can then show that when  CCEI$=0.8$,  there exists a utility function with cost efficiency $0.8$ at each observation. Now, the definition of the CCEI means that for any utility function $\tilde u$, and any $a<$20\%, there is  \textbf{one observation} $(x,p)$ (or more) in the data in which the consumer could have achieved utility $\tilde u(x)$ by spending $a$\% less than at the budget $p\cdot x$. As far as I know, this is the closest connection between the CCEI and the notion of wasting income.

In my view, the cost-efficiency interpretation does not capture the unqualified statements about wasting income that have appeared in empirical work. It is not right to say that the consumer is wasting 20\% of their income at the cost-rationalizing utility function because that would imply that the same level of utility could be achieved with 20\% less income -- actually the consumer uses all their income to achieve the cost-rationalizing utilities. It would be more useful and transparent in applied work to adopt Polisson and Quah's precise wording of the meaning of CCEI.

Two behavioral interpretations of the CCEI have been proposed.  One is through a relaxed notion of GARP ($e$-GARP): see \cite{varian1990}, \cite{halevy2018parametric}, \cite*{cherchye2020consumers}, and \cite{polisson2021rationalizability}. The other is through the work of \cite{DZIEWULSKI2020105071}, who provides a foundation, using the model of  semiorders in decision theory, to Afriat's notion of cost efficiency.  In his theory, the index reflects an inability of consumers to adequately distinguish among bundles that are similar to each other. 

I hope that the arguments against the waste interpretation are convincing. The meaning of waste in the English language, and how it is used in most of the applied literature, is not supported by the definition of the CCEI. My suggestion is to use the cost-efficiency wording in Afriat (as articulated by  \cite{polisson2021rationalizability}), $e$-GARP, or the model of  \citeauthor{DZIEWULSKI2020105071} to phrase empirical findings based on the CCEI.\footnote{\cite{halevy2018parametric} provides a useful discussion of alternative interpretations of the CCEI and other measures of distance to rationality (see page 1566).}

\subsection{The CCEI as a metric of irrationality}

The CCEI is not only used to assess whether a consumer is utility-maximizing, it is also used to compare different degrees of irrational behavior. Indeed if there are two agents, Alice and Bob, both having a CCEI substantially smaller than one, then whoever has the highest level of CCEI is said to be more rational than the other. I shall now argue that this use of CCEI is misguided. In fact, the CCEI is used to make both ordinal (``Alice is more rational than Bob'') and cardinal statements (``Alice's is twice as rational as Bob''). I am going to argue that ordinal comparisons can be problematic, and that even when they are not, the corresponding cardinal comparisons also have issues.

I present an example in which irrationality and CCEI have a non-monotonic and non-linear relation, which allows me to find behaviors in which Alice is more rational than Bob, but her CCEI is lower. In my example, irrationality stems from preference instability. The details are in Section~\ref{sec:modelandexample}, but the basic features of the example are straightforward. A consumer's behavior is guided by two selves: in odd periods, her purchases are made by her version of Mr Hyde, and in even periods by Dr Jekyll. Actually in the example there will be two observations, one generated by Mr Hyde and the other by Dr Jekyll. The difference between the two behaviors is captured by the distance between their expenditure shares --- shares that are conveniently held constant by assuming a Cobb-Douglas preference for each self. Think of Alice's behavior: In period 1 her purchases are made by her Mr Hyde preference, who has a well-behaved  Cobb-Douglas utility, and spends most of her budget on cheese and bread. In period 2, her purchases are made by her Dr Jekyll preference, also with a  Cobb-Douglas utility, who spends the most on knives and ropes. This preference instability means that Alice is irrational and violates GARP. I will measure the degree of her irrationality by the distance between her Jekyll and Hyde preferences, as measured by the distance in expenditure shares (or equivalently by the distance between the Cobb-Douglas parameters; later in the paper is an alternative example that does not depend on this choice of measure).

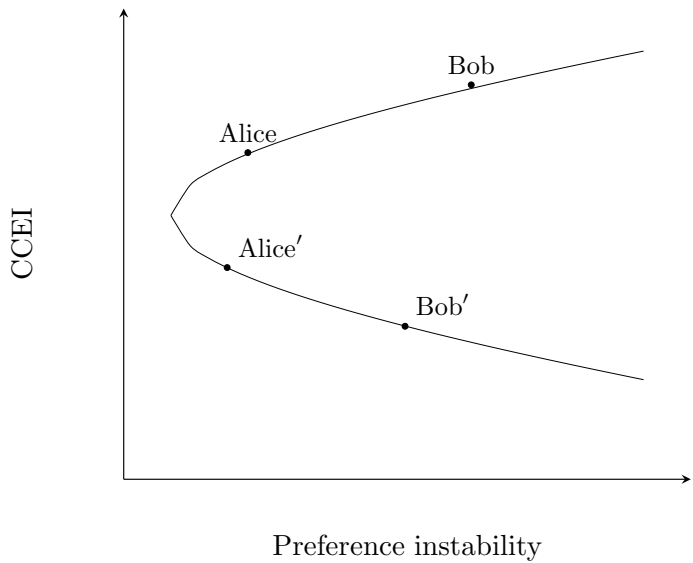
\begin{figure}[htt]
    \centering
\pgfplotsset{ticks=none}
\begin{tikzpicture}[scale=1.1]
\begin{axis}[%grid=both,
          xmax=1.4,ymax=.5,ymin=0.305,xmin=1.0001,
          axis x line=bottom,
          xlabel = {\footnotesize Preference instability},
          axis y line=left,
          ylabel = {\footnotesize CCEI},
          restrict y to domain=0:.7,
          enlargelimits]
\addplot[domain=1.000:1.4,very thin,smooth]  {5/12+(1/6)*pow(pow(x,2)/4 -1/4,1/2)} node[above]{};
\addplot[domain=1.000:1.4,very thin,smooth]  {5/12-(1/6)*pow(pow(x,2)/4 -1/4,1/2)} node[above]{};
\end{axis}
\draw[fill] (1.25,2.56) circle (1pt) node[above right] {\footnotesize $\textrm{Alice}'$};
\draw[fill] (3.4,1.85) circle (1pt) node[above right] {\footnotesize $\textrm{Bob}'$};
\draw[fill] (1.5,3.95) circle (1pt) node[above] {\footnotesize Alice};
\draw[fill] (4.2,4.77) circle (1pt) node[above] {\footnotesize Bob};
\end{tikzpicture}
    \caption{Degree of preference instability and CCEI over the parameter space in the example of Section~\ref{sec:modelandexample}. The curve depicted is defined by Equation~\ref{eq:cceisqrt} with $\da=0$. The pair of agents $(\text{Alice},\text{Bob})$, with an ``incorrect'' ordering by CCEI, results from choosing two different parameter values. The pair $(\text{Alice}',\text{Bob}')$ reflects alternative parameter values.}
    \label{fig:cceiandinstability}
\end{figure}

Figure~\ref{fig:cceiandinstability} summarizes the relation between CCEI and preference instability in the example. In particular, it is possible to choose two parameter values to obtain two agents, Alice and Bob, so that Alice is more rational than Bob but has a lower CCEI. This means that a reasonable behavioral notion of irrationality, the degree of preference instability (measured by the distance between Cobb-Douglas expenditure shares), is not reflected in the CCEI. In this sense, ordinal comparisons based on the CCEI may be misguided.

The example also serves to comment on the cardinal use of CCEI. Many papers estimate a quantitative model for observed degrees of rationality. Such a model does not only make an ordinal use of CCEI, it is also interpreted cardinally. For example, imagine a variable $X$ that takes real values, and the linear regression model $\textrm{CCEI}_i=a+b X_i$. In the linear model, changes in the values of CCEI must be explained through a corresponding change in the values of $X_i$; indeed if $X$ is measured in US dollars, then CCEI is measured in a currency that has the fixed $b$ exchange rate to the US dollar. 

Now consider the cardinal interpretation in light of the example in Figure~\ref{fig:cceiandinstability}. Even when we can restrict attention to a parametrization of the model in which CCEI and irrationality have the requisite ordinal relation (we focus on consumers like Alice' and Bob' ), the relation is non-linear. If the actual measure of irrationality is preference instability, this creates problems.

The first problem is, again, one of interpretation. A given change in CCEI cannot be interpreted in isolation. An increase of $0.05$ to the CCEI does not mean the same when we go from a value of $.5$ to $.55$ as when we go from $.8$ to $.85$. The underlying change in ``irrationality'' can be dramatically different. And, if we are comparing two different consumers, even when the ordinal comparison is valid, any cardinal comparison is meaningless.

The second problem is statistical. In Figure~\ref{fig:cceiandinstability},  the linear model is obviously misspecified, but even if we ignore this issue, any inference is plagued by bias and heteroskedasticity. If the correct econometric specification involves preference instability as the true measure of irrationality, then the nonlinear relation between the two variables implies that the variance of CCEI depends on the variance of the difference between CCEI and the measure of preference instability. It is also easy to see that nonlinearity leads to biased estimates of the relation between irrationality and other variables of interest.

\subsection{Preference instability.}

My example assumes unstable preferences as the source of irrational choice. It is the natural general model of irrational choice for consumption data. In specialized environments (choice under risk and uncertainty, or intertemporal choice, for example), there are well-documented specific sources of ``bias,'' but such models are not portable from one environment to the next. Unstable preferences are portable, and they constitute the canonical model of irrational behavior, in the sense that any data can be trivially reconciled with sufficiently unstable preferences.
Finally, unstable preferences have historically been seen as the natural source of irrationality: Samuelson viewed the weak axiom of revealed preference as a ``consistency'' condition, a reference to preference invariance; and proponents of rational choice argued against the existence of unstable preferences. 
Gary Becker and George Stigler % \cite{stigler1977gustibus} 
likened preferences to the Rocky Mountains in their permanence and invariance. Milton Friedman made similar statements. And \cite{landsburg1981taste} used revealed preference theory to argue that consumers' preferences in the UK remained stable from 1900 to 1955.

I have chosen to measure instability through the distance between Cobb-Douglas parameters. This choice is arbitrary, but reasonable given that these parameters encode the constant expenditure shares involved in each violation. In any case it is easy to come up with examples that do not rely on the distance  between agents' preferences. Here is one such example: Alice and Bob have, again, unstable preferences. They have Mr Hyde and Dr Jekyll preferences, both of which are Cobb-Douglas with respective parameters $\al$ and $\al'$: the \textit{same} for Alice and Bob, so we will not be comparing them. Alice has, however, much more stable preferences than Bob. She uses her Hyde preference in 9 out of 10 periods, while Bob's choices alternate between his Hyde and Jekyll preferences.  We may, with some confidence, claim that Alice is more rational than Bob: we might even be tempted to say that she is \textit{much more rational} because her behavior is most of the time consistent with a single preference. As we shall see, this may not be reflected in CCEI.

Concretely, suppose that there are two goods and that prices change from $p=(1/3,1)$ in odd periods to 
$p'=(1/2,1/2)$ in even periods (budgets are normalized to 1). 
Suppose that $\al=(1/10,9/10)$ and $\al'=(9/10,1/10)$. 
So Alice will choose 
$(3/10,9/10)$ in periods $1,3,5,7,9$, 
$(2/10,18/10)$ in periods $2,4,6,8$, and
$(18/10,2/10)$ in period 10. 
Bob, in contrast, will choose $(3/10,9/10)$ in odd periods at prices $p$  and $(18/10,2/10)$ in even periods at prices $p'$. It is easy to verify that the datasets generated by Alice and Bob have the same  CCEI of 0.8.  So we conclude that Bob's much larger preference instability is not reflected in a smaller CCEI.

\section{The model and example}\label{sec:modelandexample}

A \df{dataset} is a collection $\{(x^k,p^k):1\leq k \leq K \}$, where each $x^k\in\Re^n_+$ is a consumption bundle and each $p^k\in\Re^n_{++}$ a price vector. The data corresponds to the observation of a single agent, a consumer, who purchases $x^k$ at time $k$ when the prevailing prices where $p^k$, and her budget was $I^k=p^k\cdot x^k$. 

Given a dataset and a number $e\in[0,1]$, we may define the \df{$e$-revealed preference} relation as a binary relation $R(e)$ on $\{x^k:1\leq k \leq K\}$ by $x^k\mathrel{R(e)} x^l$ if and only if $ep^k\cdot x^k\geq p^k\cdot x^l$. Denote the transitive closure of $R(e)$ by $R^T(e)$. Then a dataset is $e$-acyclic if $x^l \mathrel{R^T(e)} x^k$ implies that $ep^k\cdot x^k\leq p^k\cdot x^l$. A dataset satisfies the \df{Generalized Axiom of Revealed Preference} (GARP) if it is $1$-acyclic. 

The CCEI is usually defined as the largest $e\in[0,1]$ for which the dataset is $e$-acyclic.\footnote{Strictly speaking this is not a formal definition because there may exist no such largest $e$. In practice this usually does not matter --- however see \cite{murphy2015caveat}.} Several variants of CCEI have been proposed in the literature, including Varian's index \citep{variangoodness1993}, and the money pump index \citep{echenleeshum2011}. My comments on interpretation do not apply to the money pump index, but the thrust of the example below does apply to both Varian's and the money pump indexes.

I now turn to an example with four goods and two observations, so $n=4$ and $K=2.$ A Cobb-Douglas utility with parameter vector  $\al=(\al_1,\ldots,\al_4)\geq 0$ and $\sum_l\al_l=1$ gives rise to a demand function \[ d^\al_l(p,I) =  \frac{\al_l I}{p_l}, \quad  l=1,2,3,4
\] that depends on price $p$ and income $I$.

Fix two parameters $\al$ and $\al'$. Consider a dataset $(x^k,p^k)$ in which odd observations are generated from demand $d^\al$ while even observations are generated from demand $d^{\al'}$. By suitably normalizing prices, we may without loss of generality  suppose that income is always equal to one (so $p^k\cdot x^k=1$). 

Given the numbers $\da>0$ and $\ta\in (-1/2+\da,1/2-\da)$, consider the following values for $\al$, $\al'$, $p$ and $p'$:
\[
\begin{array}{c|cccc}
l & 1 & 2 & 3 & 4 \\ \hline
\al(\ta) & \frac{1}{2}+\ta -\da & \frac{1}{2} - \ta -\da & \da & \da \\
\al'(\ta) & \da & \da & \frac{1}{2} +\ta -\da& \frac{1}{2} - \ta-\da  \\
p & 2 & 3 & 1 & 1 \\
p' & 1 & 1 & 2 & 3 \\
\end{array}
\] Throughout, $\da$ should be thought of positive but very small: the only purpose of allowing for $\da>0$ instead of $\da=0$ is to ensure that agents' preferences are strictly monotonic. A casual reader might overwise think that the behaviors documented here are incompatible with strictly monotonic preferences. It, however, easiest to read the next calculation by assuming that $\da=0$, and that is the assumption behind the drawing in Figure~\ref{fig:cceiandinstability}. Then we have that
\begin{align*}
    p'\cdot d^{\al(\ta)}(p,1) & = \sum_i \al_i(\ta) (p'_i/p_i) = \frac{1}{6}(5/2 + \ta + 25\da) \\
    p\cdot d^{\al'(\ta)}(p',1) & = \sum_i \al'_i(\ta) (p_i/p'_i) = \frac{1}{6}(5/2 + \ta + 25\da). \\
\end{align*}
Consider the data set with $(x^1,p^1)=(\da^{\al(\ta)}(p,1),p)$ and $(x^2,p^2)=(\da^{\al'(\ta)}(p',1),p')$, which then exhibits  a GARP violation for any $\ta\in (-1/2+\da,1/2-\da)$. The CCEI in this example is \[\max\{ p'\cdot d^{\al(\ta)}(p,1), p\cdot d^{\al'(\ta)}(p',1)\}= 5/12 +\ta/6 + (25/6)\da.\]

On the other hand \[ 
\norm{\al(\ta) - \al'(\ta)}^2 = 2[(1/2-\ta-2 \da)^2 + (1/2+\ta- 2 \da)^2]
= 1+4\ta^2 + 8\da(2\da-1).
\]
which is strictly increasing in $\ta$ for $\ta>0$, but strictly decreasing for $\ta<0$. So if we compare two different agents obtained by setting two different, but positive, values of $\ta$, then one will be more irrational in the sense of have more unstable preferences while seeming more rational by having a larger value of CCEI.

Further, if we set $i=\norm{\al(\ta) - \al'(\ta)}$ as a measure of irrationality, then we have that 
\begin{equation}\label{eq:cceisqrt}
\textrm{CCEI}=\frac{1}{6}\left( \frac{5}{2}  \pm \sqrt{\frac{i^2-1-8\da(2\da-1)}{4}} + 25\da
\right).
\end{equation} This is the relation that is graphed in Figure~\ref{fig:cceiandinstability}.

\subsection{Cost efficiencies and aggregation.} One way to understand the CCEI is as the result of two separate steps. The idea of a two-step procedure was suggested to me by Matt Polisson, and has been emphasized in the literature by \cite{halevy2018parametric}. First, for each observation $(x^k,p^k)$ we may consider the \df{cost efficiency coefficient} $e^k\in [0,1]$ and decide that $x^k$ is revealed preferred to $x^l$ when $e^kp^k\cdot x^k\geq p^k\cdot x^l$. The set of vectors $(e^k)$ for which the resulting revealed preference relation is acyclic (essentially the $e$-GARP idea alluded to earlier, but with a more flexible treatment of the $e$ coefficients) captures the degree of violations of rationality in the data. The second step is to aggregate the set of cost-efficiency coefficients into a single number. The CCEI results from a particular aggregation procedure (see in particular the online apendix to \cite{halevy2018parametric} for a detailed description and discussion of different aggregation procedures).

The first, pre-aggregation, step is perhaps the fundamental idea behind the CCEI. It is shared by Varian's index, and the money-pump indexes, which differ in how cost-efficiency coefficients are aggregated. So it is important to note that the preference-instability example also presents issues with the first step of the CCEI. Indeed, for the chosen parameter values, the set of ``rational'' cost-efficiency coefficients expands while preference instability increases. For the Alice and Bob consumers depicted in Figure~\ref{fig:cceiandinstability}, the set of cost efficiency coefficients that preclude a violation of GARP consists of all vectors $(e^1,e^2)$ with $e^k\leq\text{CCEI}$. So the example is not driven by the aggregation assumption in the CCEI, and its negative conclusions extends to measures that differ from in aggregation procedure.

\subsection{Robustness}
The example is obviously special, but not knife-edge in any sense. One may perturb the parameters of the model ($\al$ and $\al'$), and the conclusions I have drawn continue to hold. The same is true of prices, so my conclusions do not depend too much on picking the right prices. A simple calculation of randomly drawn prices show that the data in the example do not have small probability. For example choosing a small and a large value of $\ta$ (as for Alice and Bob in Figure~\ref{fig:cceiandinstability}, but with $\ta$ close to 0 for Alice and $\ta$ close to $1/2$ for Bob), and assuming that relative prices are drawn from a uniform distribution, the conclusion in my example, meaning the positive relation between CCEI and $i$, occurs with probability arbitrarily close to $9/12$. This particular number has of course no practical importance, but shows that the example is not in any way negligible, nor are the prices completely arbitrary.

\subsection{Budget dependence}

It is well known that CCEI depends on the particular collection of budgets faced by an agent, and researchers are usually careful in comparing the values of CCEI obtained under different experimental designs. Indeed it is easy to construct examples of the kind I have presented in this paper, where an agent has unstable preferences and presents different values of CCEI when faced with different collections of budgets. 

Perhaps more interesting is that one can find such examples in which purely random choice does not detect a difference between the budgets. This matters because it is common among practitioners to use purely random behavior to gauge the likelihood that an experimental design catches a violation of GARP. But there are situations where random behavior produces the same outcomes in two different sets of budgets, but an agent with unstable preferences exhibits different values of CCEI. 

To see this, consider all the price vectors that can be achieved through a permutation of the goods indexes -- a natural way to generate prices for an experimental design. By the symmetries inherent in the group of permutations, random choices may not detect any different between alternative subsets of prices, but a fixed preference instability (as in the Jekyll/Hyde example) will produce different results in the different resulting budgets.\footnote{A trivial example with three gods is: $p_1=(1,1,0.1)$, $p_2=(1,0.1,1)$, $p_3=p_1$ and $p_4=(0.1,1,1)$. The first choice is made by Hyde with Cobb-Douglas parameter $\al=(0.01,0.98,0.01)$, and the second by Jekyll with $\al'=(0.01,0.01,0.98)$. An experiment with income $=1$ and prices $p_1$ followed by $p_2$ with produce a CCEI of $0.2$, while $p_3$ followed by $p_4$ with produce no WARP violation. Simulated random choice will not distinguish between these two designs.}

\section{Discussion}

\subsection{Other critiques of the CCEI}
\paragraph{\textbf{Lack of benchmark.}} I have emphasized the problems in ascribing meaning to particular values of the CCEI. This issue is well know. \cite{variangoodness1993} advocated the use of numbers that are familiar from the application of statistical significance (he suggested $0.95$ for ``sentimental reasons''), but acknowledged that there was no real basis for picking these numbers. The problem is recognized and addressed by \cite{cherchye2020consumers}, who use a permutation test to construct an alternative statistical theory based on the CCEI as a test for the null or irrational random behavior. 

\paragraph{\textbf{Weakest link.}}
Another known issue with CCEI is that it tries to break revealed preference cycles at their weakest link, which can make it sensitive to extreme observations. This problems serves to motivate the indexes of \cite{varian1990} and \cite*{echenleeshum2011}. An empirical confirmation of such lack of robustness is in \cite{opatrny2018extent}, who shows that small values of CCEI are often driven by a single observation.

\paragraph{\textbf{CCEI$=1$.}}
It is also well known that the CCEI may equal one when there is in effect a violation of GARP. This can happen when one of the revealed-preference comparisons in the cycle occurs with equality. \cite{murphy2015caveat} argue empirically that such violations are not as rare as one might think.

\paragraph{\textbf{Units of measurement.}}
\cite{allen2021measuring} show that the units in which CCEI is expressed matters. Their finding is related to the problems I have highlighted regarding the cardinal use of the CCEI, but their view is that different cardinal expressions may be needed to tackle different substantive problems. 

\subsection{An approach to testing for preference instability}

Given the emphasis on preference instability in this note, I would like to outline an approach to testing for rationality that tackles preference instability directly. The idea is to provide for a utility function that explains each individual observation in a dataset, and then to require the utility functions to be close to each other. So if the data has $K$ observations, $K$ different utility functions. The dataset is \df{perfectly rationalizable} when these utility functions may be taken to be identical to each other, and \df{close to being rationalizable} when these $K$ utilities are similar to each other.

Given a dataset $\{(x^k,p^k):1\leq k\leq K \}$, 
Afriat's theorem states that the existence of a (locally nonsatiated) utility function that rationalizes the data is equivalent to the existence of a solution to a system of linear inequalities \[
V^l\leq V^k + \la^k p^k\cdot (x^l-x^k)
\] for each $1\leq k,l\leq K$. The unknowns are the scalar variables $V^k$ and $\la^k>0$. Now it is important to note that a rationalizing utility may be constructed and will achieve utility $V^k$ at observation $k$, with gradient $\la^k p^k$.

Consider an arbitrary dataset; one that may or may not be rationalizable. I introduce a new system of linear inequalities:
Fix an observation $k$. The following system of linear inequalities provides a utility function that has the ``correct gradient'' at the $k$th observation:
\begin{enumerate}
  \item 
For each $h$ and each $l\neq k$ we have an inequality:
\[
U^h_k \leq U^l_k + q^l_k \cdot (x^h-x^l),
\] in which the unknowns are the scalars $U^h_k$ and $U^l_k$, and the vector $q^l_k\in\Re^{\ngoods}_{+}$.
\item For each $l$ we have an inequality
  \[
U^l_k \leq U^k_k + \la^k p^k \cdot (x^l-x^k),
\] in which the unknowns are the scalars $U^l_k$ and $U^k_k$, and $\la^k>0$. 
\end{enumerate}
There are $K(K-1)$ inequalities of the first kind, and $K$ inequalities of the second kind. The unknowns are the $K^2$ scalar variables $U^h_k$, the $K$ scalar variables $\la^k$, and the $K(K-1)$ $n$-dimensional vectors $q^l_k$. Key here is that a solution to this system provides, through the standard Afriat construction, a utility that has the ``correct'' gradient for demanding $x^k$ at prices $p^k$, and gradient at which it will demand other bundles $x^l$ that may not coincide with what is needed for optimality at prices $p^l$. The system, however, remains linear and of manageable size.

In all, considering the systems defined by each different $k$, we obtain a linear system that has a polynomial number of inequalities and variables. Let $P$ be the set of solutions $(U_k,q_k,\la^k)\in \Re^K\times\Re^{K\ngoods}\times \Re_+$, for $1\leq k\leq K$, to the corresponding system of linear inequalities. Note that $P$ has a computationally tractable definition. Moreover:

\begin{obs}
$P$ is a non-empty convex polytope.
\end{obs}

It is easy to see that the rationalizability of the data is directly connected to the existence of ``Afriat-like'' solutions in $P$:

\begin{proposition*}
The dataset is rationalizable if and only if there is a solution with $U^h_k=U^h_l$ for all $h,l,k$ and $q^l_k = \la^l p^l$ for all $l,k$.
\end{proposition*}

The (very simple) proof of the proposition is omitted.

With the set $P$ in place, we may consider the following convex program:
\[
\Phi = \min \{\sum_{k,h} \norm{U_k-U_h} + \sum_{l,k}\norm{q^l_k - \la^lp^l} :
(U_k,q_k,\la^k)_{k=1}^K\in P
\}
\]

The proposition implies that a dataset is rationalizable if and only if $\Phi=0$. More importantly, one can build on $\Phi$ to obtain a measure of distance to rationality in the data, and a statistical test for the null hypothesis of perfect rationalizability. The details are omitted, but the methodology for building a test from the convex program defining $\Phi$ is standard.  The test would be powerful in large samples against the alternative of preference instability.

\section{Conclusion}\label{sec:conclusion}

As a test for rationality, the CCEI makes intuitive and mathematical sense, but its values are difficult to interpret. A CCEI of $0.8$ has no self-evident economic meaning, so there is no common-sense criterion to say ``this number is too low for me to conclude that the consumer is utility maximizing.'' 

For the purpose of testing rationality, there are alternative measures. The money pump index of \cite*{echenleeshum2011}, which is based on very similar ideas to the CCEI, has a simple monetary interpretation as the amount of money that could be extracted from a consumer that violates GARP. The magnitudes have an economic interpretation.\footnote{In personal communication, Pawe\l\ \citeauthor{DZIEWULSKI2020105071} has shown a model that provides a behavioral interpretation to the money pump index.} Importantly, it also provides a formal test statistic for rationality, and with some additional assumptions can be used to conduct statistical inference. One may use the money pump index to calculate $p$-values, and thus support a decision to reject the null hypothesis of rationality.%\footnote{The money pump index was motivated by the difficulties in interpreting the CCEI, and addresses other problems with the CCEI, see \cite{echenleeshum2011} for a detailed discussion.} 
Varian's index can also be used to conduct a formal statistical test (\cite{varian1990}). Another proposal is that of \cite*{cherchye2020consumers}, who takes random behavior as the null hypothesis and uses the CCEI to test against the alternative of utility maximization. \citeauthor{cherchye2020consumers} are also able to devise a statistical test.

The problem of deciding whether a dataset is rational, and measuring degrees of irrationality, is clearly interesting and important. I am not suggesting that anyone should abandon this line of research. I do suggest, as a  practical empirical procedure that is immediately available,  to use a battery of different tests, report alternative results, and highlight the pros and cons of the different measures. A few existing papers have already introduced this practice. I also hope to motivate some new theoretical work. There are exciting ideas for a principled axiomatic approach, such as \cite{apesteguia2015measure}, \cite{mononen2020foundations},  \cite{efeokandgereltcomparative}, \cite{declippel2021relaxed}, and \cite{ribeirocomparative}. It is also possible that success depends on focusing on specific choice domains.  The exercise of comparing degrees of irrationality is intrinsically difficult. There are too many ways in which people can be irrational, and it is possible that the phenomenology of irrationality is too rich to capture with a single numerical index. In specific domains there are known source of ``biases'' that can be leveraged in constructing indexes of rationality.

\clearpage
\bibliographystyle{ecta}
\bibliography{ccei}

\begin{thebibliography}{19}
\newcommand{\enquote}[1]{``#1''}
\expandafter\ifx\csname natexlab\endcsname\relax\def\natexlab#1{#1}\fi

\bibitem[\protect\citeauthoryear{Afriat}{Afriat}{1972}]{afria72}
\textsc{Afriat, S.} (1972): \enquote{Efficiency Estimation of Production
  Functions,} \emph{International Economic Review}, 13, 568--598.

\bibitem[\protect\citeauthoryear{Afriat}{Afriat}{1973}]{afriat1973system}
---\hspace{-.1pt}---\hspace{-.1pt}--- (1973): \enquote{{On a system of
  inequalities in demand analysis: an extension of the classical method},}
  \emph{International Economic Review}, 14, 460--472.

\bibitem[\protect\citeauthoryear{Allen and Rehbeck}{Allen and
  Rehbeck}{2021}]{allen2021measuring}
\textsc{Allen, R. and J.~Rehbeck} (2021): \enquote{Measuring rationality:
  Percentages vs expenditures,} \emph{Theory and Decision}, 1--13.

\bibitem[\protect\citeauthoryear{Apesteguia and Ballester}{Apesteguia and
  Ballester}{2015}]{apesteguia2015measure}
\textsc{Apesteguia, J. and M.~A. Ballester} (2015): \enquote{A measure of
  rationality and welfare,} \emph{Journal of Political Economy}, 123,
  1278--1310.

\bibitem[\protect\citeauthoryear{Chambers and Echenique}{Chambers and
  Echenique}{2016}]{chambers2016revealed}
\textsc{Chambers, C.~P. and F.~Echenique} (2016): \emph{Revealed Preference
  Theory}, Cambridge University Press (Econometric Society Monographs).

\bibitem[\protect\citeauthoryear{Cherchye, Demuynck, De~Rock, and
  Lanier}{Cherchye et~al.}{2020}]{cherchye2020consumers}
\textsc{Cherchye, L., T.~Demuynck, B.~De~Rock, and J.~Lanier} (2020):
  \enquote{Are consumers rational? Shifting the burden of proof,} FEB Research
  Report Department of Economics, KU Leuven.

\bibitem[\protect\citeauthoryear{{de Clippel} and Rozen}{{de Clippel} and
  Rozen}{2021}]{declippel2021relaxed}
\textsc{{de Clippel}, G. and K.~Rozen} (2021): \enquote{Relaxed Optimization:
  How Close is a Consumer to Satisfying First-Order Conditions?} Forthcoming,
  Review of Economics and Statistics.

\bibitem[\protect\citeauthoryear{Dziewulski}{Dziewulski}{2020}]{DZIEWULSKI2020105071}
\textsc{Dziewulski, P.} (2020): \enquote{Just-noticeable difference as a
  behavioural foundation of the critical cost-efficiency index,} \emph{Journal
  of Economic Theory}, 188, 105071.

\bibitem[\protect\citeauthoryear{Echenique, Lee, and Shum}{Echenique
  et~al.}{2011}]{echenleeshum2011}
\textsc{Echenique, F., S.~Lee, and M.~Shum} (2011): \enquote{The Money Pump as
  a Measure of Revealed Preference Violations,} \emph{Journal of Political
  Economy}, 119, 1201--1223.

\bibitem[\protect\citeauthoryear{Halevy, Persitz, and Zrill}{Halevy
  et~al.}{2018}]{halevy2018parametric}
\textsc{Halevy, Y., D.~Persitz, and L.~Zrill} (2018): \enquote{Parametric
  recoverability of preferences,} \emph{Journal of Political Economy}, 126,
  1558--1593.

\bibitem[\protect\citeauthoryear{Landsburg}{Landsburg}{1981}]{landsburg1981taste}
\textsc{Landsburg, S.~E.} (1981): \enquote{Taste change in the United Kingdom,
  1900-1955,} \emph{Journal of Political Economy}, 89, 92--104.

\bibitem[\protect\citeauthoryear{Mononen}{Mononen}{2020}]{mononen2020foundations}
\textsc{Mononen, L.} (2020): \enquote{Foundations of measuring (cardinal)
  rationality,} Mimeo, University of Z\"urich.

\bibitem[\protect\citeauthoryear{Murphy and Banerjee}{Murphy and
  Banerjee}{2015}]{murphy2015caveat}
\textsc{Murphy, J.~H. and S.~Banerjee} (2015): \enquote{A caveat for the
  application of the critical cost efficiency index in induced budget
  experiments,} \emph{Experimental Economics}, 18, 356--365.

\bibitem[\protect\citeauthoryear{Ok and Tserenjigmid}{Ok and
  Tserenjigmid}{2021}]{efeokandgereltcomparative}
\textsc{Ok, E. and G.~Tserenjigmid} (2021): \enquote{Comparative rationality of
  random choice behavior,} Unpublished Working Paper.

\bibitem[\protect\citeauthoryear{Opatrny}{Opatrny}{2018}]{opatrny2018extent}
\textsc{Opatrny, M.} (2018): \enquote{Extent of irrationality of the consumer:
  Combining the Critical Cost Eciency and Houtman Maks Indices,} .

\bibitem[\protect\citeauthoryear{Polisson and Quah}{Polisson and
  Quah}{2021}]{polisson2021rationalizability}
\textsc{Polisson, M. and J.~Quah} (2021): \enquote{Rationalizability,
  Cost-Rationalizability, and Afriat’s Efficiency Index,} .

\bibitem[\protect\citeauthoryear{Ribeiro}{Ribeiro}{2020}]{ribeirocomparative}
\textsc{Ribeiro, M.} (2020): \enquote{Comparative rationality,} Unpublished
  Working Paper.

\bibitem[\protect\citeauthoryear{Varian}{Varian}{1990}]{varian1990}
\textsc{Varian, H.~R.} (1990): \enquote{Goodness-of-fit in optimizing models,}
  \emph{Journal of Econometrics}, 46, 125 -- 140.

\bibitem[\protect\citeauthoryear{Varian}{Varian}{1993}]{variangoodness1993}
---\hspace{-.1pt}---\hspace{-.1pt}--- (1993): \enquote{Goodness-of-fit for
  revealed preference tests,} Working Paper University of Michigan.

\end{thebibliography}

\end{document}